\documentclass[aps,prl,twocolumn,groupedaddress,showpacs,floatfix]{revtex4}
\usepackage{dcolumn}
\usepackage{amsmath}
\usepackage{amssymb}
\usepackage{graphicx}
\usepackage{bm}
\usepackage[T1]{fontenc}
\usepackage{color}
\usepackage{url}
\usepackage[bf]{subfigure}
\usepackage{rotating}

\begin{document}

\title{Cluster Explosive Synchronization in Complex Networks}

\author{Peng Ji$^{1,2}$}
\email{pengji@pik-potsdam.de}
\author{Thomas K.DM. Peron$^3$}
\email{thomas.peron@usp.br}
\author{Peter J. Menck $^{1,2}$}
\author{Francisco A. Rodrigues$^4$} 
\email{francisco@icmc.usp.br}
\author{J\"urgen Kurths$^{1,2,5}$}
\affiliation{$^1$Potsdam Institute for Climate Impact Research (PIK), 14473 Potsdam, Germany\\
$^2$Department of Physics, Humboldt University, 12489 Berlin, Germany\\
$^3$Instituto de F\'{\i}sicade S\~{a}o Carlos, Universidade de S\~{a}o Paulo, Av. TrabalhadorS\~{a}o Carlense 400, Caixa Postal 369, CEP 13560-970, S\~{a}o
Carlos, S\~ao Paulo, Brazil\\
$^4$Departamento de Matem\'{a}tica Aplicada e Estat\'{i}stica, Instituto de Ci\^{e}ncias Matem\'{a}ticas e de Computa\c{c}\~{a}o,Universidade de S\~{a}o Paulo, Caixa Postal 668,13560-970 S\~{a}o Carlos,  S\~ao Paulo, Brazil\\
$^5$Institute for Complex Systems and Mathematical Biology, University of Aberdeen, Aberdeen AB24 3UE, United Kingdom}

\begin{abstract}
The emergence of explosive synchronization has been reported as an abrupt transition in complex networks of first-order Kuramoto oscillators. 
In this Letter, we demonstrate that the nodes in a second-order Kuramoto model, perform a cascade of transitions
toward a synchronous macroscopic state,
which is a novel phenomenon that we call \textit{cluster explosive synchronization}. 
We provide a rigorous analytical treatment using a mean-field analysis in uncorrelated networks. 
Our findings are in good agreement with numerical simulations
and fundamentally deepen the understanding of microscopic mechanisms toward synchronization.

\end{abstract}

\pacs{89.75.Hc,89.75.Kd,05.45.Xt}

\maketitle


In the past few years, much research effort has been devoted to investigating the influence of network organization on dynamical processes, 
such as random walks~\cite{noh2004random}, congestion~\cite{boccaletti2006complex,barrat2008dynamical}, 
epidemic spreading~\cite{barrat2008dynamical} and synchronization~\cite{Pecora98:PRL,arenas2008synchronization}. 
Regarding synchronization of coupled oscillators, it has been demonstrated that the emergence of collective behavior
 in these structures depends on the patterns of connectivity of the underlying network. For instance, through a mean-field analysis, it has been found that Kuramoto oscillators display 
a second-order phase transition to the synchronous state with a critical coupling strength that depends on the network topology~\cite{arenas2008synchronization}.


Recently, discontinuous transitions to phase synchronization have been observed 
in SF networks~\cite{PhysRevLett.106.128701,PhysRevLett.108.168702,PhysRevE.86.016102,PhysRevE.86.056108}. 
This phenomenon, called explosive synchronization, was proved to be caused exclusively by a microscopic correlation between the network 
topology and the intrinsic dynamics of each oscillator. 
More specifically, G\'{o}mez-Garde\~{n}ez \textit{et al}.~\cite{PhysRevLett.106.128701} considered the natural frequencies positively correlated with 
the degree distribution of the network, 
defining the natural frequency of each oscillator as equal to its number of connections.

In this Letter, we substantially extend a first-order Kuramoto model used in~\cite{PhysRevLett.106.128701} to 
a second-order Kuramoto model~\cite{PhysRevLett.109.064101,PhysRevLett.81.2229,doi:10.1137/110851584,PhysRevE.71.016215} 
that we modify in order to analyze global synchronization, 
considering the natural frequency of each node proportional to its degree~\cite{Microgrids_John_Florian,Pluchino2006a,PhysRevE.61.6987}. 
In this model, we find a discontinuous phase transition 
in which small degree nodes join the synchronous component simultaneously, 
whereas other nodes synchronize successively according to their degrees 
(in contrast to~\cite{PhysRevLett.106.128701} 
where all the nodes join the synchronous component abruptly):
This is a novel phenomenon which we call \textit{cluster explosive synchronization}. 
By developing a mean-field theory we derive self-consistent equations 
that produce the lower and upper critical coupling strength associated to 
a hysteretic behavior of the synchronization for uncorrelated networks. 
The analytical results are in good agreement with numerical simulations.
 Moreover, we show that decreasing the network average frequency and increasing the coupling strength are 
the key factors that lead to cluster explosive synchronization.

The second-order Kuramoto model consists of the following set of equations~\cite{PhysRevLett.109.064101,PhysRevLett.81.2229,doi:10.1137/110851584,PhysRevE.71.016215}:
\begin{equation}
 \frac{d^{2}\theta_{i}}{dt^{2}}=-\alpha\frac{d\theta_{i}}{dt}+\Omega_i+\sum_{j=1}^{N}\lambda_{ij}A_{ij}\sin(\theta_{j}-\theta_{i})
\label{Eq:kuramoto}
 \end{equation}
where $\theta_i$ is the phase of the oscillator $i=1,\ldots,N$, $\alpha$ the dissipation parameter, 
$\Omega_i$ the the natural frequency distributed with a given probability density $g(\Omega)$, $\lambda_{ij}$ the coupling strength 
and $A_{ij}$ an element of the network's adjacency matrix $\mathbf{A}$.
Here we assume that all 
connections have the same coupling which leads to a homogeneous coupling strength $\lambda_{ij}=\lambda$, $\forall$ $i,j$.
To study the influence of dynamics and structure on global synchronization, we assume
\begin{equation}
\Omega_i=D(k_i-\left\langle k \right\rangle)
\label{Eq:P}
\end{equation}
where $k_i$ is the degree of node $i$, $\left\langle k \right\rangle$ the network average degree and $D$ a proportionality constant. 
A mean-field analysis allows us to investigate the dynamics of the model.

We follow the continuum limit approach proposed in~\cite{Ichinomiya04:PRE} 
and assume zero degree correlation between the nodes in the network.
Denote by $\rho{(k;\theta,t)}$ the fraction of nodes of degree $k$ that have  
phase $\theta$ at time t. 
The distribution $\rho{(k;\theta,t)}$ is normalized according to 
$\int_0^{2\pi}\rho{(k;\theta,t)} d\theta =1$.
In an uncorrelated network, the probability that a randomly picked edge is 
connected to a node with degree $k$, phase $\theta$ at time $t$ is 
given by $kP(k)\rho{(k;\theta,t)}/\left\langle k \right\rangle$, 
where $P(k)$ is the degree distribution and $\left\langle k \right\rangle$ the
 average degree. 
For a network, the order parameter $r$ is defined as the magnitude of 
$r e^{i \psi(t)} = \sum_i k_i e^{i\theta_i(t)}/\sum_i k_i$.
In the continuum limit this yields
\begin{equation}
    re^{i\psi}=\int_{k_{min}}^\infty dk\int_0^{2 \pi}d\theta P(k)k/{\left\langle k \right\rangle}\rho(k;\theta,t)e^{i\theta(t)},
\label{eq:r}
\end{equation}
where $k_{min}$ is the network's minimum degree and $\psi$ is the average phase. 
When the phases $\theta$ are randomly distributed, $r\approx 0$. On the other hand,
 when the oscillators evolve with similar phases, $r\approx 1$.
 
Seeking to write a continuum-limit version of Eq.~\ref{Eq:kuramoto} with the natural frequency $D(k-\left\langle k \right\rangle)$ and the constant $\lambda$ 
in terms of the mean-field quantities $r$ and $\psi$, we multiply both sides of
Eq.~\ref{eq:r} by $e^{-i \theta}$ and take the imaginary part, obtaining
\begin{equation}
\ddot{\theta} = -\alpha \dot{\theta} + D(k-\left\langle k \right\rangle) + k\lambda r \sin(\psi - \theta),
\label{eq:kuramoto_mean_field}
\end{equation}
which is the same equation that describes the movement of a damped driven pendulum. 

In order to derive sufficient conditions for synchronization, we choose a reference frame 
that rotates with the average phase $\psi$ of the system, defining
$\phi(t) = \theta(t) - \psi(t)$.
Substituting the transformed variables $\phi(t)$ into the equations of motion 
Eq.~\ref{eq:kuramoto_mean_field} and 
defining $C(\lambda r)\equiv (\ddot{\psi}+\alpha \dot{\psi})/D$, we get
 \begin{equation}
\ddot{\phi}=-\alpha\dot{\phi}+D(k-\left\langle k \right\rangle-C(\lambda r))- k\lambda r\sin \phi.
\label{Eq:motion_continuum_phi}
\end{equation}

Eq.~\ref{Eq:motion_continuum_phi} can be interpreted as an extension to the 
second-order case of the model recently proposed 
in~\cite{PhysRevLett.106.128701}.
The first-order Kuramoto model studied in~\cite{PhysRevLett.106.128701} 
presents hysteresis only when SF topologies are 
considered in which each node's natural frequency is proportional to its degree.
In contrast, it is known that systems described by a second-order Kuramoto model present 
hysteresis independently of the choice of the natural 
frequency distribution~\cite{tanaka1997self,tanaka1997first}.

In 
order to obtain sufficient conditions for the existence of the synchronous solution of Eq.~\ref{eq:r}, we derive a 
self-consistent equation for the order parameter $r$ that
can be written as the sum of the contribution 
$r_{lock}$ due to oscillators which are phase-locked to the mean-field 
and the contribution of the non-locked drift oscillators $r_{drift}$, 
i.e., $r = r_{lock} + r_{drift}$ \cite{tanaka1997self}.

Locked oscillators are characterized by $\dot{\phi} = \ddot{\phi} = 0$, 
and turn out to possess degrees in a certain range $k \in [k_1, k_2]$. Each locked oscillator has a $k$-dependent constant phase,
$\phi = \arcsin{(\frac{\left|D\left(k - \left\langle k \right\rangle - C(\lambda r)\right)\right|}{k\lambda r})}$, 
i.e.,~$\rho(k; \phi,t)$ is a time-independent single-peaked distribution.
For the locked oscillators we obtain
\begin{equation}
r_{lock} = \frac{1}{\left\langle k \right\rangle}\int_{k_1}^{k_2}kP(k)\sqrt{1-\left(\frac{D\left(k-\left\langle k\right\rangle -C(\lambda r)\right)}{k\lambda r}\right)^2}. 
\label{Eq:int_r_lock}
\end{equation}

On the other hand, the phase of the drift oscillators rotates with period $T$ in the stationary state, 
so that their density
$\rho(k; \phi,t)$ satisfies $\rho \sim |\dot{\phi}|^{-1}$~\cite{tanaka1997self}. As
$\oint \rho(k;\phi) d\phi =\int^{T}_0 \rho(k;\phi) \dot\phi dt =1 $, this implies
$\rho(k;\phi)=T^{-1}|\dot\phi^{-1}|$~\cite{tanaka1997self}.
After substituting $\rho(k;\phi)$ into Eq.~\ref{eq:r} and performing some mathematical manipulations motivated by~\cite{tanaka1997self}, we get
\begin{equation}
r_{drift} = \left(-\int_{k_{min}}^{k_1} + \int_{k_2}^{\infty} \right)\frac{-rk^2\lambda \alpha^4 P(k)}{D^3\left(k - C(\lambda r)-\left\langle k \right\rangle\right)^3\left\langle k\right\rangle} dk.
\label{Eq:int_r_drift}
\end{equation}
The order parameter $r$ is determined by the sum of Eq.~\ref{Eq:int_r_lock} and~\ref{Eq:int_r_drift}.

It is known that systems subject to the equations of motion
 given by Eq.~\ref{eq:kuramoto_mean_field} present a hysteresis as $\lambda$ is varied~\cite{strogatz1994nonlinear,tanaka1997self}. 
Therefore we consider in the following the system's dynamics for two distinct cases: 
(i) Increasing the coupling strength $\lambda$.
In this case, the system starts without synchrony ($r\approx 0$) and, as $\lambda$ is increased, approaches the synchronous state ($r\approx 1$). 
(ii) Decreasing the coupling strength $\lambda$.
Now the system starts at the synchronous state ($r\approx 1$) and, as $\lambda$ is decreased, 
ever more oscillators lose synchronism, falling into the drift state. 
\begin{figure}[!tpb]
 \centerline{\includegraphics[width=1\linewidth]{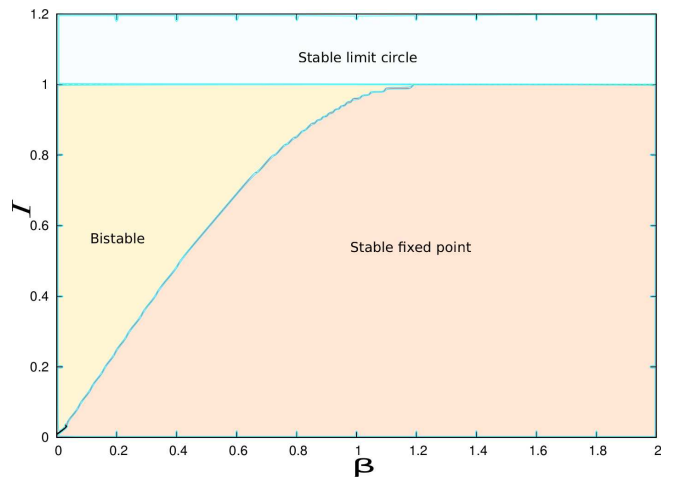}}
  \caption{(Color online) Parameter space of the pendulum (Eq. ~\ref{Eq:kuramoto_mean_field_new_time_scale}). Plotted is the external constant torque $I$ versus 
the damping strength $\beta$. The red area indicates parameter combinations that give rise to a globally stable fixed point. 
In the white area, only a stable limit cycle exists. Yellow indicates the area of bi-stability: both fixed point and limit cycle are stable.
}
  \label{Fig:Parameter_space}
\end{figure}

For the case in which the coupling strength $\lambda$ is increased from $\lambda_0$, the synchronous state emerges after a threshold $\lambda^I_c$ 
has been crossed.
 Here we derive self-consistent equations that allow to compute $\lambda^I_c$. In order to do that, we first investigate how the range of degree $k$
of oscillators that are entrained in the mean-field depends on $\lambda$. For convenience, we change the
time-scale in Eq.~\ref{Eq:motion_continuum_phi} to $\tau = \sqrt{k\lambda r}t$, which yields
\begin{equation}
\frac{d^2\phi}{d^2\tau} + \beta \frac{d\phi}{d\tau} + \sin \phi = I,
\label{Eq:kuramoto_mean_field_new_time_scale}
\end{equation}
where we define $\beta \equiv \alpha / \sqrt{\lambda k r}$ and
 $I \equiv D(k - \left\langle k \right\rangle - C(\lambda r))/k\lambda r$. The variable  
$\beta$ is the damping strength and $I$ corresponds to a constant torque 
(cf. the damped driven pendulum~\cite{strogatz1994nonlinear,guckenheimer1983nonlinear}). Our change of time-scale allows us to employ
Melnikov's analysis~\cite{guckenheimer1983nonlinear} to determine
the range of integration $\left[k_1^{I}, k_2^{I} \right]$ in the calculation of $r^I = r_{lock}^{I} + r_{drift}^I$.

Using Melnikov's analysis~\cite{tanaka1997self,guckenheimer1983nonlinear,strogatz1994nonlinear} we find that, for $I>1$, Eq.~\ref{Eq:kuramoto_mean_field_new_time_scale} 
has only one stable limit cycle solution (see Fig.~\ref{Fig:Parameter_space}). If $4\beta/\pi \leq I \leq 1$, the system is bistable
 and a synchronized state co-exists 
with the limit cycle solution. If the coupling strength is increased further by a small amount, 
the synchronized state can only exist for $I\leq 4\beta / \pi$, where 
Eq.~\ref{Eq:kuramoto_mean_field_new_time_scale} has a stable fixed point solution, even for damping values close
 to $\beta \simeq 1$~\cite{tanaka1997self,guckenheimer1983nonlinear,strogatz1994nonlinear}. 
By solving the inequalities $\sin\phi \leq 1$ and $I \leq 4\beta / \pi$, 
we get the following range of $k^I$ for the phase-locked oscillators
\begin{eqnarray}\nonumber
    k^I \in [k^I_1, k^I_2] &\equiv&
    \left[\frac{B-\sqrt{B^{2}-4D^{4}\left(\left\langle k\right\rangle +C(\lambda r)\right)^{2}}}{2D^{2}}\right.,\\
    &  & \hspace{-3ex} \left.\frac{B+\sqrt{B^{2}-4D^{4}\left(\left\langle k\right\rangle +C(\lambda r)\right)^{2}}}{2D^{2}}\right]
 \label{Eq:range_k_increase}
\end{eqnarray}
where
\begin{equation}\nonumber
B = 2D^2(\left\langle k \right\rangle + C(\lambda r)) + \frac{16\alpha^2 \lambda r}{\pi^2}.
\label{Eq:constant_B}
\end{equation}
If we now substitute Eq.~\ref{Eq:range_k_increase} into the self-consistent Eqs.~\ref{Eq:int_r_lock} 
and ~\ref{Eq:int_r_drift}, we obtain $r^I$ and $\lambda^I_c$.
\begin{figure}[tpb]
\centerline{\includegraphics[width=1\linewidth]{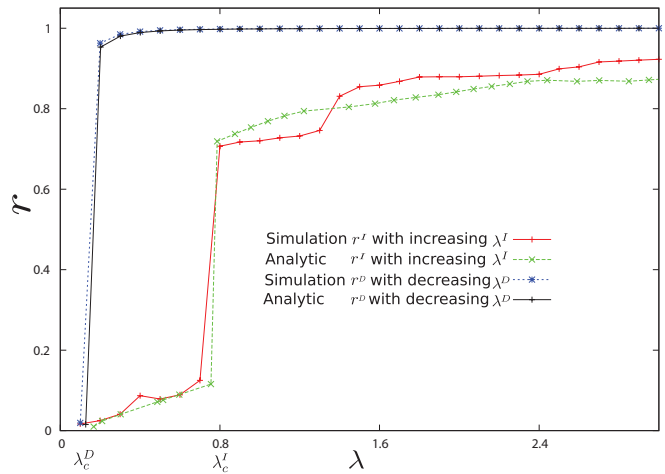}}
  \caption{(Color online) Comparison between numerical and analytical results. Shown is the order parameter for increasing (decreasing) $\lambda$ based on: 
i) simulations, red (blue) line and ii) self-consistent equations (~\ref{Eq:int_r_drift},~\ref{Eq:int_r_lock}) 
with synchronized degree $k^I$in Eq.~\ref{Eq:range_k_decrease}($k^D$ in Eq.~\ref{Eq:range_k_increase}) , green (black) line. 
}
\label{Fig:forward_backward_coupling}
\end{figure}

When the coupling strength $\lambda$ is decreased, the oscillators 
start at the phase-locked synchronous state, reaching the asynchronous state after a threshold 
$\lambda_c^{D}$. 
In order to calculate the threshold, we again investigate the range of degree $k$ of phase-locked oscillators.
Imposing the phase locked solution 
in Eq.~\ref{Eq:motion_continuum_phi}, we obtain $\sin\phi = \frac{\left|D\left(k - \left\langle k \right\rangle - C(\lambda r)\right)\right|}{k\lambda r}\leq 1$ 
and find that the locked oscillators are the nodes with degree $k$ in the following range as a function of $\lambda r$
\begin{equation}
    k^D \in [k_1^D, k_2^D] \equiv \left[\frac{\left\langle k \right\rangle + C(\lambda r)}{1 + \frac{\lambda r}{D}} , \frac{\left\langle k \right\rangle + C(\lambda r)}{1 - \frac{\lambda r}{D}} \right].
\label{Eq:range_k_decrease}
\end{equation}
This allows us to calculate  $r^D$ and $\lambda^D_c$ from the self-consistent Eqs.~\ref{Eq:int_r_lock} and ~\ref{Eq:int_r_drift}.

To check the validity of our mean-field analysis, we conduct numerical 
simulations of the model with $\alpha=0.1$ and $D=0.1$ on SF networks characterized by
$N=3000$, $\left \langle k \right \rangle = 10$, $k_{min}=5$ and the degree distribution $P(k) \sim k^{-\gamma}$ with $\gamma=3$.
Again, because we anticipate hysteresis, we have to distinguish two cases: first, we gradually increase $\lambda$ from $\lambda_0$ by amounts of $\delta \lambda$, and for $\lambda =
\lambda_0,\lambda_0 + \delta \lambda,...,\lambda_0+n\delta \lambda$ compute the increasing order parameter $r^I$, where $\delta \lambda = 0.1$.
Second, we gradually decrease $\lambda$ from $\lambda_0+n\delta \lambda$ back toward $\lambda_0$ by amounts of $\delta \lambda$, 
this time computing the decreasing order parameter $r^D$. Before each $\delta \lambda$-step, we simulate the system long enough to arrive at an attractor.

According to Fig.~\ref{Fig:forward_backward_coupling}, the dependence of the simulated order parameter $r$ on $\lambda$ indeed shows the expected hysteresis: 
for increasing $\lambda$, the discontinuous transition to synchrony happens at a $\lambda^I_c$ that is far larger than the threshold $\lambda_c^D$ of 
the backwards transition for decreasing $\lambda$. Seeking to compare these simulation results to our mean-field analysis,
 we simultaneously solve Eqs. (\ref{Eq:int_r_lock}, \ref{Eq:int_r_drift}, \ref{Eq:range_k_increase}) 
(resp. Eqs. (\ref{Eq:int_r_lock},  \ref{Eq:int_r_drift}, \ref{Eq:range_k_decrease})) for the increasing (resp. decreasing) 
case numerically to obtain analytical curves of $r$. 
\begin{figure}[tpb]
\centerline{\includegraphics[width=1\linewidth]{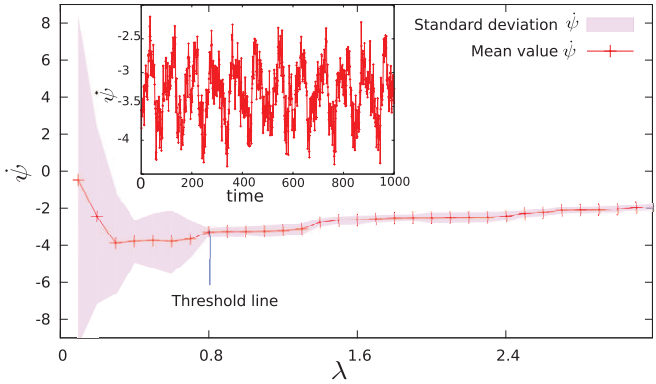}}
\caption{(Color online) Results from the numerical simulations with increasing coupling strength. The red line shows the mean value of $\dot{\psi}$($\lambda$), 
the red shade its standard deviation. The inset contains a periodic time series of $\dot\psi$ at coupling strength 0.8.}
\label{Fig:derivative_psi}
\end{figure}

A key ingredient of the solution process is $C(\lambda r)$ which we retrieve from the simulation data. 
More specifically, recalling that $C(\lambda r)$ just depends on $\dot{\psi}$ and $\ddot{\psi}$, we assume that (i) $C(\lambda r) \approx 0$ 
before the transition to synchrony, as the nodes oscillate independently and produce an unchanging zero mean field,
 and that (ii) $C(\lambda r) \approx \alpha \dot{\psi}/D$ after the transition to synchrony, as the synchronized nodes dominate and 
produce a mean field that rotates with a constant frequency $\dot{\psi}$ (which we take from the simulations,
 cf. Fig.~\ref{Fig:derivative_psi}). Fig.~\ref{Fig:forward_backward_coupling} reveals that our mean-field analysis predicts the critical thresholds $\lambda_c^I$ 
and $\lambda_c^D$ very well.
\begin{figure}[tpb]
\centerline{\includegraphics[width=1\linewidth]{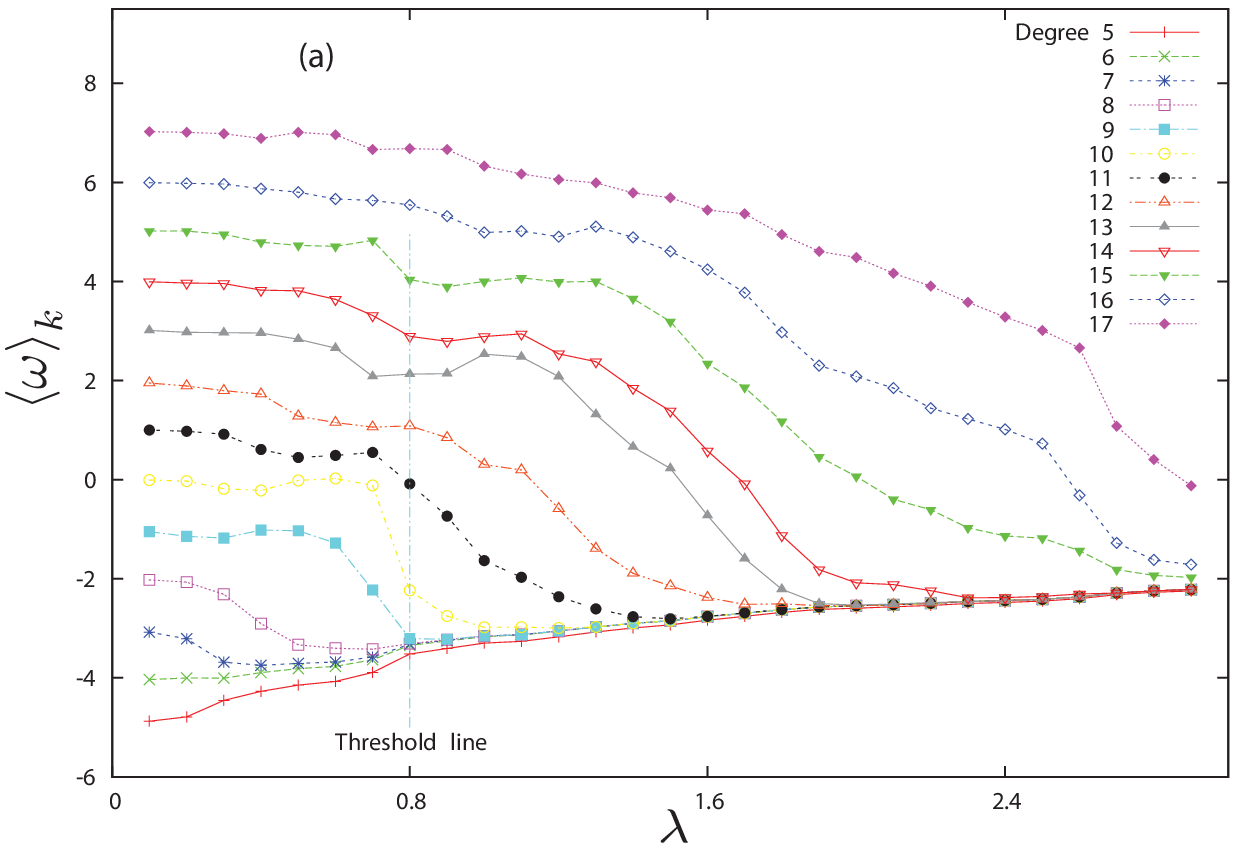}}
\centerline{\includegraphics[width=1\linewidth]{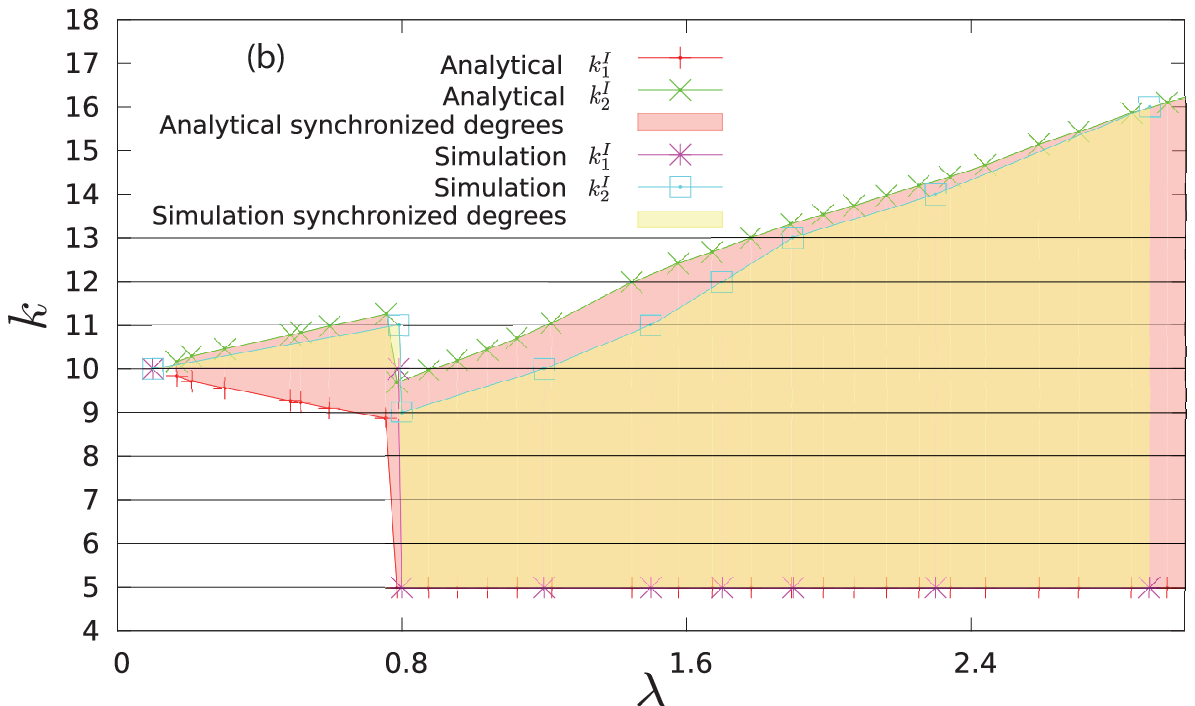}}
\caption{(Color online) (a) Results from numerical simulations with increasing coupling strength $\lambda$. Shown is the average frequency of nodes of various degrees against $\lambda$. 
(b) synchronized degrees from simulations and mean field analysis.}
  \label{Fig:Synchronized_degree_coupling}
\end{figure}

What happens at the node level when the transition to synchrony occurs?
As just stated, the average frequency $\dot{\psi}$ that enters the equations via $C(\lambda r)$ 
is a crucial quantity. Seeking to understand how
the different nodes contribute to $\dot{\psi}$, for increasing $\lambda$ 
we calculate the average frequency of all nodes of degree $k$,
$\left \langle \omega \right \rangle_k = \sum_{[i|k_i=k]} {\omega_i}/(NP(k))$, where 
$\omega_i=\int^{t+T}_t\dot\phi_i(\tau)dt / T$.
As 
Fig.~\ref{Fig:Synchronized_degree_coupling}(a) reveals, 
nodes of the same degree form clusters that join the synchronous component successively, starting from small degrees.
This is in sharp contrast to the discontinuous phase transition observed in 
\cite{PhysRevLett.106.128701}, where the average frequency 
$\left \langle \omega \right \rangle_k$ jumps
to $\left\langle k \right\rangle$ at $\lambda_c^I$ for all $k$ at the same time. 
We call this newly observed phenomenon \textit{cluster explosive synchronization}.

In Fig.~\ref{Fig:Synchronized_degree_coupling}(b), 
we show the evolution of the lower and upper limits of the range of synchronous degrees,
 $k^I_1$ and $k^I_2$, as a function of the coupling strength $\lambda$.
Analytical and simulation results are again in good agreement. 
 Note the discontinuity in 
the evolution of $k^I_1$ and $k^I_2$ which gives rise to the discontinuous transition 
in Fig.~\ref{Fig:forward_backward_coupling}. 
For $\lambda > \lambda_c^I$, the lower limit  $k^I_1$ is kept constant, 
and the higher limit $k^I_2$ of the synchronous nodes grows linearly 
with $\lambda$ (Fig.~\ref{Fig:Synchronized_degree_coupling}(b)).
%

In summary, we have demonstrated that a \textit{cluster explosive synchronization} transition occurs in a second-order Kuramoto model. 
As in previous studies on explosive synchronization, 
a correlation between dynamics (natural frequency of a node) and local topology (the node's degree) 
is a necessary condition.
With the connection between natural frequencies and local topology, we have presented the first analytical treatment of cluster explosive synchronization 
which is based on a mean-field approach in uncorrelated complex networks. Our simulations are in good agreement with the theory.
Furthermore, we have shown that clusters of nodes of the same degree join the synchronous component successively, starting with small degrees.
Our findings enhance the understanding of \textit{cluster explosive synchronization} in the macrostate of a system 
and its applications will have a strong impact on the detection of clusters in larger networks. 
Also, our first analytical treatment can be extended to applications~\cite{PhysRevLett.109.064101,PhysRevLett.81.2229,doi:10.1137/110851584,PhysRevE.71.016215}
where the use of second-order Kuramoto oscillators is relevant.

P. Ji would like to acknowledge China Scholarship Council (CSC) scholarship. 
T. Peron would like to acknowledge FAPESP.
F. A. Rodrigues would like to acknowledge CNPq (305940/2010-4) and FAPESP (2010/19440-2) for the financial support given to this research. 
P.J.~Menck acknowledges support from Konrad-Adenauer-Stiftung. 
J. Kurths would like to acknowledge IRTG 1740 (DFG and FAPESP) for the sponsorship provided.

\bibliographystyle{apsrev}
\bibliography{paper}

\end{document}